# Recognizing well-parenthesized expressions in the streaming model

Frédéric Magniez[*]        Claire Mathieu[†]        Ashwin Nayak[‡]

## Abstract

Motivated by a concrete problem and with the goal of understanding the sense in which the complexity of streaming algorithms is related to the complexity of formal languages, we investigate the problem DYCK($s$) of checking matching parentheses, with $s$ different types of parenthesis.

We present a one-pass randomized streaming algorithm for DYCK(2) with space O($\sqrt{n}\log n$), time per letter polylog($n$), and one-sided error. We prove that this one-pass algorithm is optimal, up to a polylog $n$ factor, even when two-sided error is allowed, and conjecture that a similar bound holds for any constant number of passes over the input.

Surprisingly, the space requirement shrinks drastically if we have access to the input stream in *reverse*. We present a two-pass randomized streaming algorithm for DYCK(2) with space O(($\log n$)$^2$), time polylog($n$) and one-sided error, where the second pass is in the reverse direction. Both algorithms can be extended to DYCK($s$) since this problem is reducible to DYCK(2) for a suitable notion of reduction in the streaming model. Except for an extra O($\sqrt{\log s}$) multiplicative overhead in the space required in the one-pass algorithm, the resource requirements are of the same order.

For the lower bound, we exhibit hard instances ASCENSION($m$) of DYCK(2) with length $\Theta(mn)$. We embed these in what we call a "one-pass" communication problem with $2m$-players, where $m = \tilde{O}(n)$. To establish the hardness of ASCENSION($m$), we prove a direct sum result by following the "information cost" approach, but with a few twists. Indeed, we play a subtle game between public and private coins for MOUNTAIN, which corresponds to a primitive instance ASCENSION(1). This mixture between public and private coins for MOUNTAIN results from a balancing act between the direct sum result and a combinatorial lower bound for MOUNTAIN.

[*]LRI, Univ. Paris-Sud, CNRS; F-91405 Orsay, France; magniez@lri.fr. Supported in part by the French ANR Defis program under contract ANR-08-EMER-012 (QRAC project), and the French ANR Sesur program under contract ANR-07-SESU-013 (VERAP project)

[†]Computer Science Department, Brown University; Providence RI 02912, U.S.A; claire@cs.brown.edu. Part of this work was funded by NSF grant CCF-0728816.

[‡]C&O and IQC, U. Waterloo and Perimeter Institute; 200 University Ave. W. Waterloo, ON N2L 3G1, Canada; anayak@math.uwaterloo.ca. Work conducted at Rutgers University and DIMACS Center while on sabbatical leave from Waterloo. Research supported in part by NSERC Canada. Research at Perimeter Institute is supported in part by the Government of Canada through Industry Canada and by the Province of Ontario through MRI.

# 1 Introduction

The area of streaming algorithms has experienced tremendous growth over the last decade in many applications. Streaming algorithms sequentially scan the whole input piece by piece in one pass, or in a small number of passes (i.e., they do not have random access to the input), while using sublinear memory space, ideally polylogarithmic in the size of the input. The design of streaming algorithms is motivated by the explosion in the size of the data that algorithms are called upon to process in everyday real-time applications, for example in bioinformatics for genome decoding, in Web databases for the search of documents, or in network monitoring. The analysis of Internet traffic [2], in which traffic logs are queried, was one of the first applications of this kind of algorithm. Few applications have been made in the context of formal languages, which may have impact on massive data such as DNA sequences and large XML files. For instance, in the context of databases, properties decidable by streaming algorithm have been studied [24, 23], but only in the restricted case of deterministic and constant memory space algorithms.

Motivated by a concrete problem and with the goal of understanding the sense in which the complexity of streaming algorithms is related to the complexity of formal languages, we investigate the problem $\textsc{Dyck}(s)$ of checking matching parentheses, with $s$ different types of parenthesis. Regular languages are by nature decidable by deterministic streaming algorithms with constant space. The $\textsc{Dyck}$ languages are some of the simplest context-free languages and yet already powerful. The $\textsc{Dyck}(s)$ language plays a central role in context-free languages, since every context-free language $L$ can be mapped to a subset of $\textsc{Dyck}(s)$ [9]. In addition to its theoretical importance, the problem of checking matching parentheses is enountered frequently in database applications, for instance in verifying that an XML file is well-formed.

Deciding membership in $\textsc{Dyck}(s)$ has already been addressed in the massive data setting, more precisely through property testing algorithms. An $\varepsilon$-property tester [6, 7, 12] for a language $L$ accepts all strings of $L$ and rejects all strings which are $\varepsilon$-far from strings in $L$, for the normalized Hamming distance. For every fixed $\varepsilon > 0$, $\textsc{Dyck}(1)$ is $\varepsilon$-testable in constant time [1], whereas in general $\textsc{Dyck}(s)$ is $\varepsilon$-testable in time $\tilde{O}(n^{2/3})$, with a lower bound of $\tilde{\Omega}(n^{1/11})$ [21]. In [11], a comparison between property testers and streaming algorithms has been made. One advantage of streaming algorithms is that they have access to the full string, albeit not in a random access fashion.

With random access to the input, context-free languages are known to be recognizable in space $O((\log n)^2)$ [13]. In the special case of $\textsc{Dyck}(s)$, logarithmic space is sufficient, as we may run through all possible heights, and check parentheses at the same height. This scheme does not seem to easily translate to streaming algorithms, even with a small number of passes over the input.

In the streaming model, $\textsc{Dyck}(1)$ has a one-pass streaming algorithm with logarithmic space, using a height counter. Using a one-way communication complexity argument for $\textsc{Equality}$, we can show that $\textsc{Dyck}(2)$ requires linear space for deterministic one-pass streaming algorithms. A relaxation of $\textsc{Dyck}(s)$ is $\textsc{Identity}(s)$ in the free group with $s$ generators, where local simplifications $\bar{a}a = \epsilon$ are allowed in addition to $a\bar{a} = \epsilon$, for every type of parenthesis $(a, \bar{a})$. There is a logarithmic space algorithm for recognizing the language $\textsc{Identity}(s)$ [18] that can easily be massaged into a one-pass streaming algorithm with polylogarithmic space. Again, this algorithm does not extend to $\textsc{Dyck}(s)$.

We show that $\textsc{Dyck}(s)$ is reducible to $\textsc{Dyck}(2)$, for a suitable notion of reduction in the streaming model, with a $\log s$ factor expansion in the input length. Our first algorithm is a one-pass randomized streaming algorithm for $\textsc{Dyck}(2)$ with space $O(\sqrt{n}\log n)$ and time $\text{polylog}(n)$ (**Theorem 1**). If we had no space constraints the algorithm would be very simple: when we encounter an upstep ($a$ or $b$), push it on a stack, when we encounter a downstep ($\bar{a}$ or $\bar{b}$), pop the top item from the stack and check that they match. However the stack may grow to linear size in this process. To avoid this growth, the basic principle of our algorithm is that instead, we use a linear hash function to periodically (every $\sqrt{n}$ letters) compress the information. As long as we compress only upsteps or only downsteps, all at different heights, we are able to detect mismatches with high probability. The algorithm has one-sided error; it accepts words that belong to the language with certainty. Although simple, we show that this appealing algorithm is nearly optimal in its space usage, even when two-sided error is allowed (**Corollary 1**).

We conjecture that our lower bound still holds if we read the stream several times, but always in the same direction. Surprisingly, the situation is drastically different if we can read the stream in *reverse*. We present a second algorithm, a randomized two-pass streaming algorithm for $\textsc{Dyck}(2)$ with $\mathrm{O}((\log n)^2)$ space and time $\mathrm{polylog}(n)$, where the second pass is in the reverse direction (**Theorem 2**). This algorithm uses a hierarchical decomposition of the stream into blocks; whenever the algorithm reaches the end of a block, it compresses the information about subwords within the block. This compression is what reduces the stack size from $\sqrt{n}$ down to $\mathrm{O}(\log n)$, but prevents us from checking that certain pairs of parentheses are well-formed. However, given the profile of the word (i.e., the sequence of heights), we can pinpoint exactly the pairs that do not get checked. As it turns out, a pair that does not get checked when scanning the input left to right will necessarily be checked when scanning in the reverse direction. Like the one-pass algorithm, this algorithm has only one-sided error, and always accepts words that belong to the language. We note that it is easy to extend the algorithms so that it recognizes the language of substrings (which are subsequences of consecutive letters) of $\textsc{Dyck}(2)$.

As mentioned above, we also investigate the lower bound on the space required for any one-pass randomized streaming algorithms. Such a lower bound is by nature hard to prove because of the connection of the problem with $\textsc{Identity}(2)$. Moreover, proving any lower bound based on two-party communication complexity is hopeless: the related communication problem automatically reduces to $\textsc{Equality}$ after local checks and simplifications by both players, thereby proving only an $\Omega(\log n)$ lower bound. Instead, we build hard instances $\textsc{Ascension}(m)$ of $\textsc{Dyck}(2)$ with length $\Theta(mn)$, that we embed in a "one-pass" communication problem with $2m$ players, where $m = \tilde{\Theta}(n)$. The constraint is that the length of each message in the protocol be less than size, a function of $n$. Our main result (**Theorem 4**) is that such a protocol requires size $= \Omega(n)$, which proves that our one-pass algorithm is nearly optimal (**Corollary 1**).

To establish the hardness of $\textsc{Ascension}(m)$, we prove a *direct sum* result that captures its relationship to solving $m$ instances of the intermediate problem $\textsc{Mountain}$, which involves only two players. We follow the "information cost" approach taken in [8, 22, 4, 17, 15], among other works before and since. We adapt this notion to suit both the nature of streaming algorithms and of our problem. The idea is to focus on the information about a part of the input contained in a part of the protocol transcript, given the remaining inputs.

Using this notion of information cost, we prove a direct sum result (**Lemma 3**). A remarkable device here is the use of an "easy" distribution for the information cost for protocols, that are correct with high probability in the worst case. The use of an easy distribution "collapses" $\textsc{Ascension}(m)$ to an instance of $\textsc{Mountain}$, which may be planted in any one of the $m$ coordinates. This technique was developed in [4], but comes with a few twists in our case. Indeed, we play a subtle game between public and private coins. Namely, in protocols for $\textsc{Ascension}(m)$ only public coins are allowed for all players, whereas for $\textsc{Mountain}$ one of the players, Bob, can also access private coins, while Alice, the other player, cannot. This mixture between public and private coins for $\textsc{Mountain}$ arises from a balancing act between the direct sum result and our combinatorial lower bound for $\textsc{Mountain}$ (**Theorem 3**). Namely, we are only able to prove the lower bound for $\textsc{Mountain}$ when Alice only uses public coins, whereas the direct sum only holds, with our definition of information cost, when Bob has access to additional private coins.

We note that as a bonus, our lower bound provides a $\tilde{\Omega}(\sqrt{n})$ lower bound for the problem of checking priority queues in the one-pass streaming model, solving an open problem of [10].

## 2 Definitions and preliminaries

Here is a formal definition of the language of parentheses with $s$ types of parenthesis.

**Definition 1** (DYCK). *Let $s \geq 1$ be an integer and let $\Sigma = \{a_1, \overline{a}_1, \ldots, a_s, \overline{a}_s\}$. Let $\mathrm{DYCK}(s)$ be the language over $\Sigma$ defined recursively by $\mathrm{DYCK}(s) = \epsilon + a_1\mathrm{DYCK}(s)\overline{a}_1 + \ldots + a_s\mathrm{DYCK}(s)\overline{a}_s$.*

We also denote by $\mathrm{DYCK}(s)$ the problem of deciding if $w$ is in the language $\mathrm{DYCK}(s)$, given the word $w \in \Sigma^*$.

We recall the notion of streaming algorithms, where one *pass* on an input $x \in \Sigma^n$ means that $x$ is given as an *input stream* $x_1, x_2, \ldots, x_n$, which arrives sequentially, i.e., letter by letter in this order. For simplicity, we assume throughout this article that the length $n$ of the input is always given to the algorithm in advance. Nonetheless, all our algorithms can be adapted to the general case where $n$ is unknown until the end of a pass. For an excellent introduction to streaming algorithms, we refer the reader to the book [19].

**Definition 2** (Streaming algorithm). *Fix an alphabet $\Sigma$. A $k$-pass streaming algorithm $A$ with space $s(n)$ and time $t(n)$ is an algorithm such that for every input stream $x \in \Sigma^n$: (1) $A$ performs $k$ sequential passes on $x$; (2) $A$ maintains a memory space of size $s(n)$ while reading $x$; (3) $A$ has running time at most $t(n)$ per letter $x_i$; (4) $A$ has preprocessing and postprocessing time $t(n)$.*
*We say that $A$ is* bidirectional *if it is allowed to access to the input in the reverse order, after reaching the end of the input. Then the parameter $k$ is the total number of passes in either direction.*

**Definition 3** (Streaming reduction). *Fix two alphabets $\Sigma_1$ and $\Sigma_2$. A problem $P_1$ is $f(n)$-streaming reducible to a problem $P_2$ with space $s(n)$ and time $t(n)$, if for every input $x \in \Sigma_1^n$, there exists $y = y_1 y_2 \ldots y_n$, with $y_i \in \Sigma_2^{f(n)}$, such that: (1) $y_i$ can be computed deterministically from $x_i$ using space $s(n)$ and time $t(n)$; (2) From a solution of $P_2$ with input $y$, a solution on $P_1$ with input $x$ can be computed with space $s(n)$ and time $t(n)$.*

**Fact 1.** *Let $P_1$ be $f(n)$-streaming reducible to a problem $P_2$ with space $s_0(n)$ and time $t_0(n)$. Let $A$ be a $k$-pass streaming algorithm for $P_2$ with space $s(n)$ and time $t(n)$. Then there is a $k$-pass streaming algorithm for $P_1$ with space $s(n \times f(n)) + s_0(n)$ and time $t(n \times f(n)) + t_0(n)$ with the same properties as $A$ (deterministic/randomized, unidirectional/bidirectional).*

**Proposition 1.** $\mathrm{DYCK}(s)$ *is $\lceil \log s \rceil$-streaming reducible to $\mathrm{DYCK}(2)$ with space and time $\mathrm{O}(\log s)$.*

*Proof.* We encode a parenthesis $a_i$ by a word of length $l = \lceil \log s \rceil$ with only parenthesis $a_1, a_2$ as follows. We let $f(a_i)$ be the binary expansion of $i$ over $l$ bits where $0$ is replaced by $a_1$ and $1$ by $a_2$. Then $f(\overline{a}_i)$ is defined similarly, except that we write the binary expansion of $i$ in the opposite order. Then $x_1 \ldots x_n$ is in $\mathrm{DYCK}(s)$ if and only $f(x_1) \ldots f(x_n)$ is in $\mathrm{DYCK}(2)$. □

Note that in most often cases, such as XML files, the above reduction can be implemented within constant space an time. Indeed, given an upstate (*start-tag*) $<w>$ (resp. an downstep (*end-tag*) $</w>$), where $w$ is an ASCII string denoting the type of the parenthesis (*tag*), we can generate the above encoding of $w$ into $a_1, a_2$ (respectively, into $\overline{a}_1, \overline{a}_2$), while reading $w$ as a stream itself, i.e., character by character.

By Proposition 1, it is enough to design streaming algorithms for $\mathrm{DYCK}(2)$. That is the objective of the next section.

## 3 Algorithms

Throughout this section we consider $\textsc{Dyck}(2)$ where the input is a stream of $n$ letters $x_1 x_2 \ldots x_n$ in the alphabet $\Sigma = \{a, \bar{a}, b, \bar{b}\}$. We first introduce a few definitions.

**Definition 4** (Height, Matching pair, Well-formed)**.** *Let $x \in \Sigma^n$.*
*An* upstep *(respectively,* downstep*) is a letter $a$ or a $b$ (respectively, $\bar{a}$ or $\bar{b}$).*
*The* height *of $x$ is* $\mathrm{height}(x) = |x|_a + |x|_b - |x|_{\bar{a}} - |x|_{\bar{b}}$.
*For $1 \le i < j \le n$, the pair $(i, j)$ is a* matching pair *for $x$ if* $\mathrm{height}(x[1, i]) = \mathrm{height}(x[1, j-1])$ *and* $\mathrm{height}(x[1, k]) > \mathrm{height}(x[1, i])$ *for all $k \in \{i + 1, \ldots, j - 2\}$.*
*The* height *of a matching pair $(i, j)$ is* $\mathrm{height}(x[1, i])$.
*A matching pair $(i, j)$ for $x$ is* well-formed*, if $(x[i], x[j])$ equals $(a, \bar{a})$ or $(b, \bar{b})$.*

These definitions are extended to subsets $I \subseteq [1, n]$ of indices of letters of $x$. For instance, we say that $I$ is a *matching set* for $x$, if $I = \cup\{i, j\}$, where the union is over a subset of the matching pairs $(i, j)$ for $x$. For ease of notation, we identify an increasing sequence $i_1 < i_2 < \cdots < i_m$ of indices with the corresponding sub-word $x_{i_1}, x_{i_2}, \ldots, x_{i_m}$ of $x$. We also use this correspondence in reverse when the indices of the sub-word are clear from the context. Last we need a weaker condition than well-formedness:

**Definition 5.** *Let $x \in \Sigma^n$, $I \subseteq [n]$ and $l \in \{a, b\}$. Then $I$ is $l$-balanced if for all $d \ge 0$,*

$$|\{i \in I \ : \ x_i = l, \mathrm{height}(x[1, i]) = d\}| \quad = \quad |\{i \in I \ : \ x_i = \bar{l}, \mathrm{height}(x[1, i-1]) = d\}|.$$

*Moreover $I$ is* balanced *if it is both $a$-balanced and $b$-balanced.*

We now give a well-known characterization of $\textsc{Dyck}(2)$.

**Fact 2.** *Let $x \in \Sigma^n$. Then $x \in \textsc{Dyck}(2)$ if and only if: the height of every prefix of $x$ is nonnegative,* $\mathrm{height}(x) = 0$, *and $[1, n]$ is a well-formed set for $x$.*

During the computation the algorithm implicitly keeps track of the height of the word read so far. Let $p$ be a prime number such that $n^{1+c} \le p < 2n^{1+c}$, for some fixed constant $c \ge 1$. We assume that the algorithm has access to a random function $\mathsf{hash}$ mapping subsequences of $x$ to integers in $[0, p - 1]$, as follows: $\mathsf{hash}(x_{i_1} x_{i_2} \ldots x_{i_m}) = \sum_j \mathsf{hash}(x_{i_j})$, with

$$\mathsf{hash}(x_i) = \begin{cases} \alpha^{\mathrm{height}(x[1, i])} \bmod p & \text{if } x_i = a, \\ -\alpha^{\mathrm{height}(x[1, i-1])} \bmod p & \text{if } x_i = \bar{a}, \\ 0 & \text{otherwise,} \end{cases}$$

where $\alpha$ is a uniformly random integer in $[0, p - 1]$.

The value of $\mathsf{hash}(x)$ is a polynomial in $\alpha$ of degree bounded by the maximum height of a prefix, which is at most $n$. Therefore if $h$ is not identically zero, by Schwartz's lemma, for a random $\alpha$ the probability that $\mathsf{hash}(x) = 0$ is at most $n/p \le n^{-c}$. Therefore we get the following characterization of balanced subsequences:

**Fact 3.** *Let $x \in \Sigma^n$, and let $v = x_{i_1} x_{i_2} \ldots$ be a subsequence of letters of $x$. If $v$ is a balanced set for $x$, then* $\mathsf{hash}(v) = 0$ *for all $\alpha$; otherwise* $\mathsf{hash}(v) = 0$ *with probability at most $n^{-c}$, for a uniformly random $\alpha$.*

For any letter $v_i$, we may compute $\mathsf{hash}(v_i)$ in time $\mathrm{polylog}\, n$ and space $\mathrm{O}(\log n)$. Moreover, for any word $v$ the value of $\mathsf{hash}(v)$ may be maintained with $\mathrm{O}(\log n)$ space.

## 3.1 The one-pass algorithm

The algorithm is easiest to understand if $x = x^{(u)}x^{(d)}$, where $x^{(u)}$ has only upsteps and $x^{(d)}$ has only downsteps, in equal numbers. To check $x^{(u)}x^{(d)} \in \mathrm{DYCK}(2)$, the naive algorithm would grow a stack of size $n/2$. Here is a simple alternative. We read the input in blocks of length $\sqrt{n}$. While our algorithm is reading letters of $x^{(u)}$, the stack stores the values of $\mathrm{hash}(x[i\sqrt{n}+1,(i+1)\sqrt{n}])$ for each $i \in \{1,\dots,\sqrt{n}/2\}$ and notes that $\mathrm{height}(x[i\sqrt{n}+1,(i+1)\sqrt{n}]) = \sqrt{n}$. While the algorithm is reading $x^{(d)}$, it adds $\mathrm{hash}(x[j\sqrt{n}+1,(j+1)\sqrt{n}])$ to $\mathrm{hash}(x[i\sqrt{n}+1,(i+1)\sqrt{n}])$ for $j = \sqrt{n}-i-1$, and checks if their sum is 0. The input $x$ is ill-formed if any of the sums is non-zero. Our algorithm is a generalization of this stack compression idea.

For any downstep $x_j$, our algorithm, given $(h,\ell) = (\mathrm{hash}(v),\mathrm{height}(v))$, can easily compute $\mathrm{hash}(vx_j) = h + \mathrm{hash}(x_j)$ and $\mathrm{height}(vx_j) = \ell - 1$, without explicit knowledge of $v$. Note that this relies on the linearity of our hash function.

**Algorithm 1** reads the stream in blocks of $\sqrt{n}$ letters. It uses a stack data structure encoding the prefix formed by all the letters seen so far but whose matching pairs have not yet been checked.

---

**Algorithm 1** One-pass algorithm

$S \leftarrow$ empty stack
**for** $i \leftarrow 1$ to $\sqrt{n}$ **do**
   **Algorithm 2** with $S$     {which consumes $\sqrt{n}$ letters}
**end for**
if $S$ not empty, **reject:** "missing closing parenthesis"
**return accept**

---

For clarity of exposition, we describe an "off-line" version of **Algorithm 2** that processes the letters in a block of size $\sqrt{n}$ after the entire block has been read. With little additional effort, it can be converted to an "online" algorithm that takes $\mathrm{polylog}\, n$ time per letter.

Within a block, **Algorithm 2** reads the letters one by one, doing the obvious checks (with the straightforward algorithm that uses a linear-size stack). After simplifying the pairs that have been checked, the block is reduced to a word in $(\bar{a}+\bar{b})^*(a+b)^*$, consisting of a sequence $w'$ of only downsteps followed by a sequence $w''$ of only upsteps. To retain needed information about the blocks that have already been scanned, the algorithm uses a stack data structure. Each stack item is a pair of the form $(h,\ell)$ encoding a subsequence $v$ of letters of the stream $x$ such that $h = \mathrm{hash}(v)$ and $\ell = \mathrm{height}(v)$. Recall that the computation of $\mathrm{hash}(v)$ depends on the height of its starting point within $x$.

As the algorithm processes the letters in $w'$, it incorporates information about them into the last stack item, until the associated subsequence has height 0. At that point, to test whether $v$ is well-formed, the algorithm checks whether $\mathrm{hash}(v) = 0$. If this test succeeds, the entry of the stack encoding $v$ can now be removed. The subword $w''$ is processed in a straightforward manner by creating a new stack item associated with $w''$. An example execution of the algorithm is shown in Appendix A.

---

**Algorithm 2** One-pass subroutine: reading one block

---

{uses an input stack $S$}

read the word $w$ consisting of the next $\sqrt{n}$ letters (or less if the stream becomes empty)

check that matching pairs within $w$ are well-formed (if not, **reject:** "mismatched parentheses")

simplify $w$ into $w'w''$, where $w'$ has only downsteps and $w''$ has only upsteps

**for** $i \leftarrow 1$ to $|w'|$ **do**

  pop $(h, \ell)$ from $S$ (if empty, **reject:** "extra closing parenthesis")

  {$(h, \ell)$ encodes some subsequence $v$: $h = \text{hash}(v)$ and $\ell = \text{height}(v)$}

  $\ell \leftarrow \ell - 1$

  $h \leftarrow h + \text{hash}(w'_i)$

  push $(h, \ell)$ on $S$

  {$(h, \ell)$ now encodes $vw'_i$}

  **if** $\ell = 0$ **then**

    check that $h = 0$ (if not, **reject:** "mismatched parentheses")

    pop and discard $(h, \ell)$

  **end if**

**end for**

push $(\text{hash}(w''), |w''|)$ on $S$

{$(\text{hash}(w''), |w''|)$ encodes $w''$}

**return** the stack $S$

---

We first start with the following observation about **Algorithm 1**.

Define an order between words by taking the transitive closure of $uv \prec ul\overline{l'}v$, where $l, l' \in \{a, b\}$, i.e. $w \prec x$ if $w$ is a subsequence of $x$ obtained by removing some (well-formed or not) matching pairs in $x$.

**Fact 4.** *Consider the stack right after a new push of an item encoding a subsequence ending with $x_j$. Let $v_1, v_2, \ldots, v_m$ be the subsequences of letters of $x$ encoded by the current stack items (bottom-up order). Then every $v_i$ has positive height, and $v_1 v_2 \ldots v_m \preceq x[1, j]$.*

The above fact can be used to prove the following useful invariant of **Algorithm 1**.

**Fact 5.** *Let $(h, \ell)$ be a stack item encoding some subsequence $v$ of $x$. Then $h = \text{hash}(v)$ and $\ell = \text{height}(v) \geq 0$, and $v = v_u v_d$, where $v_u$ has only upsteps, $v_d$ has only downsteps. For all $j \in v_d$ there is a unique $i \in v_u$ such that $(i, j)$ is a matching pair for $x$. Moreover $\ell = 0$ holds only for an item on top of the stack.*

Then, we conclude with the correctness of our algorithm.

**Theorem 1.** *Algorithm 1 is a one-pass randomized streaming algorithm for $\text{DYCK}(2)$ with space $O(\sqrt{n} \log n)$ and time $\text{polylog}(n)$. If the stream belongs to $\text{DYCK}(2)$ then the algorithm accepts it with probability 1; otherwise it rejects it with probability at least $1 - n^{-c}$.*

*Proof.* In terms of space requirements, each stack element takes space $O(\log n)$ and each execution of Algorithm 2 adds at most one element to the stack, for a total of at most $\sqrt{n}$ stack items, hence space $O(\sqrt{n} \log n)$. The processing time is easy by inspection.

To prove correctness, first assume that $x \in \text{DYCK}(2)$. By Fact 2 the height of prefixes are all non-negative, so the algorithm does not reject because of an extra closing parenthesis; and the height of $x$ is 0, so the algorithm doesn't reject because of a missing closing parenthesis. For each block $w$, the matching pairs within $w$ are all well-formed, so the algorithm doesn't reject them either. Finally, whenever the algorithm checks $h = 0$ for a stack item such that $\ell = 0$, by Fact 5 the corresponding encoded subsequence $v$ is a matching set for $x$ since $x \in \text{DYCK}(2)$, so $v$ is balanced. Then by Fact 3, it passes the hash test in **Algorithm 2**. Therefore the algorithm is correct in this case.

Second, assume that $x \notin \text{DYCK}(2)$. By Fact 2, $x$ fails to be in $\text{DYCK}(2)$ for one of the following reasons. Either some prefix of $x$ has negative height (too many closing parentheses): then the algorithm detects the problem when it tries to pop an item from an empty stack, hence is correct. Or, the final height of $x$ is non-zero: then the algorithm detects the problem at the very end when it sees that the stack is not empty, hence it is correct. Or, there is a matching pair $(i, j)$ where $x$ is not well-formed: that is the only non trivial case. If $i, j$ are within the same block, then the algorithm rejects during the internal checks within the block. Assume now that $i$ and $j$ are in different blocks, and that the algorithm accepts $x$. Since the stack is empty at the end of the algorithm, at some point the stack item whose subsequence contains $x_j$ gets discarded. Let $v$ be the subsequence encoded by the stack item at that point. By Fact 5, $v$ also contains $x_i$ and, since it is unbalanced at that height, from Fact 3, the probability that $v$ passes the hash test in **Algorithm 2** is at most $n^{-c}$, for a random uniform choice of $\alpha$, so the algorithm is correct with probability $1 - n^{-c}$. $\qquad\square$

## 3.2 The two-pass algorithm

The second algorithm depends on a parameter $k = \lceil \log n \rceil$, where $n$ is the size of the input word $w$. We assume that without lost of generality, $n = 2^k$. We achieve this by padding: we append to $w$ the word $(a\bar{a})^i$ of suitable length (assuming that $w$ is of even size, otherwise $w \notin \text{DYCK}(2)$). This requires that we to store, after the first pass, the number of letters $2i$ we added. This uses only $\text{O}(\log n)$ bits of memory This assumption is crucial for the analysis, since the algorithm uses a hierarchical decomposition of the stream into nested blocks and the assumption guarantees the same decomposition, whether we read it from left to right or from right to left.

An important implicit convention we make, is that during the right to left pass, letters $\bar{a}, \bar{b}$ are resp. interpreted as $a, b$ (and vice-versa).

As before, we use a stack data structure. Each stack item contains values $h$ that have been obtained by summing $\text{hash}(x_j)$ for some $j$'s in a subsequence $v$, along with auxiliary information $\ell = \text{height}(v)$. In addition, we append to each stack item the index of the first letter in $v$, denoted $\text{first}(v)$.

---

**Algorithm 3** Bi-directional algorithm
$S \leftarrow$ empty stack
**Algorithm 4** with parameters $(k, S)$, reading the stream from left to right $\{\ k = \lceil \log n \rceil\ \}$
if $S$ is not empty, **reject:** "missing closing parenthesis"
**Algorithm 4** with parameters $(k, S)$, reading the stream from right to left
{In the right to left pass, letters $\bar{a}, \bar{b}$ are resp. interpreted as $a, b$ (and vice-versa)}
**return accept**

---

Algorithm 4 recursively decomposes the stream into blocks (Figure 2 in Appendix A). An $i$-block is a substring of the stream of the form $x[(q-1)2^i + 1, q2^i]$ for $1 \le q \le n/2^i$. The main difference between **Algorithm 4**$(k, S)$ and **Algorithm 2**$(S)$ is that whenever the algorithm reaches the end of a block, it compresses *without checking* the stack items encoding subwords from within the block. This compression is what reduces the stack size from $\sqrt{n}$ down to $\text{O}(\log n)$, but we pay for that in terms of accuracy. Since hash is commutative we lose information. For example compressing $\text{hash}(a a \bar{b})$ with $\text{hash}(b b \bar{b} \bar{a})$ gives $\text{hash}(a a \bar{b} b b \bar{b} \bar{a}) = \text{hash}(a a \bar{a} b b \bar{b} \bar{b})$: one word is ill-formed, the other one is well-formed, but after compress-ing we can no longer distinguish between them. The crux of the analysis is that such critical information loss cannot occur both when reading the stream from left to right and when reading it from right to left (Fact 7 below, and Figure 3 in Appendix A).

---

**Algorithm 4** Block algorithm $(i, \text{stack } S)$ { reads $2^i$ letters in block $B_i$, increases stack size by at most 1 }

---

**for** $j \leftarrow 1$ to $2$ **do**
    **if** $i > 1$ **then**
        {read recursively two $(i-1)$-blocks $B$ and $B'$}
        **Algorithm 4**$(i-1, S)$
    **else**
        read one letter $y$
        **if** $y$ is an upstep **then**
            push$(\text{hash}(y), 1, \text{first}(y))$ on $S$
            {$(\text{hash}(y), 1, \text{first}(y))$ encodes $y$}
        **else**
            pop $(h, \ell, f)$ from $S$ (if empty, **reject:** "negative height")
            {$(h, \ell, f)$ encodes some subsequence $v$: $h = \text{hash}(v)$, $\ell = \text{height}(v)$, $f = \text{first}(v)$}
            $\ell \leftarrow \ell - 1$ and $h \leftarrow h + \text{hash}(y)$
            push $(h, \ell, f)$ on $S$
            {$(h, \ell, f)$ now encodes $vy$}
            **if** $\ell = 0$ **then**
                check that $h = 0$ (if not, **reject:** "mismatched parentheses")
                pop and discard $(h, \ell, f)$
            **end if**
        **end if**
    **end if**
**end for**
**if** $S$ has, at the top, two items whose first letters are in $B_i$ **then**
    {there are at most two such items $v_1, v_2$, that moreover are contiguous at the top}
    pop $(h_2, \ell_2, f_2)$ from $S$ and then pop $(h_1, \ell_1, f_1)$    {$f_1, f_2$ are in $B_i$}
    {$(h_1, \ell_1, f_1)$ encodes $v_1$, $(h_2, \ell_2, f_2)$ encodes $v_2$, and each index in $v_1$ is smaller than all the ones in $v_2$}
    compress: push $(h_1 + h_2, \ell_1 + \ell_2, f_1)$ on $S$
**end if**

---

First, observe that Fact 4 remains valid for **Algorithm 3**. Using this fact, we derive the following two invariants of **Algorithm 3** that are similar to Fact 5, but more involved.

**Fact 6.** *Let* $(h, \ell, f)$ *be a stack item encoding some subsequence* $v$ *of* $x$*. Then* $h = \text{hash}(v)$*,* $\ell = \text{height}(v) \geq 0$ *and* $f = \text{first}(v)$*. For every downstep* $j \in v$ *there is an upstep* $i \in v$ *such that* $(i, j)$ *is a matching pair for* $x$*. Moreover* $\ell = 0$ *only holds for an item at the top of the stack.*

**Fact 7.** *For every* $j \in [n]$*, after reading* $x_j$ *and completing its processing, each stack item* $(h, \ell, f)$ *with* $f < j$ *satisfies* $\text{height}(x[1, f]) < \text{height}(x[1, j])$ *if* $x_j$ *is an upstep, and* $\text{height}(x[1, f]) < \text{height}(x[1, j-1])$ *if* $x_j$ *is a downstep.*

We now state a simple observation from the definition of matching pairs.

**Fact 8.** *Let $u, u', v$ be subsequences of $x$ such that $v = uu'$. Then, for every possible height $d$, there is at most one matching pair in $u \times u'$ at height $d$.*

We conclude with the correctness of our algorithm.

**Theorem 2.** *Algorithm 3 is a bidirectional two-pass randomized streaming algorithm for $\textsc{Dyck}(2)$ with space $O((\log n)^2)$ and time $\mathrm{polylog}(n)$. If the input belongs to $\textsc{Dyck}(2)$ then the algorithm accepts it with probability $1$; otherwise it rejects it with probability at least $1 - n^{-c}$.*

*Proof.* In terms of space requirements, each stack element takes space $O(\log n)$ and the stack has size at most $2k = 2 \log n$, hence space $O((\log n)^2)$. The processing time is easy by inspection, while noticing by induction that each execution of **Algorithm 4** generates only one new stack item.

To analyze the algorithm, observe (using Fact 6) that it is correct whenever $x \in \textsc{Dyck}(2)$.

Now, assume that $x \notin \textsc{Dyck}(2)$. By Fact 2, either some prefix of $x$ has negative height: then as before the algorithm is correct. Or, the final height of $x$ is non-zero: then as before the algorithm is correct. Or, there is a matching pair $(j, j')$ at some height $d$ where $x$ is not well-formed. Consider that case. Let $i$ be such that $x_j$ and $x_{j'}$ are in different $(i-1)$-blocks $B$ and $B'$ but in the same $i$-block $B_i$. Assume further that among badly formed pairs, $(j, j')$ has been chosen so that $i$ is minimum (see Figure 3 in Appendix A). Let $m$ be the minimum of $\mathrm{height}(x[1, l])$, where $l$ ranges over $B$ such that $x_l$ is an upstep. Similarly, let $m'$ be the minimum of $\mathrm{height}(x[1, l-1])$, where $l$ ranges over $B$ but now such that $x_l$ is a downstep.

Without loss of generality (up to reversing left-to-right and right-to-left directions) assume that $m \geq m'$. Indeed if $m < m'$, let $\overline{x}$ denote the reverse string $x$, where letters $\overline{a}, \overline{b}$ are resp. interpreted as $a, b$ (and vice-versa). Then $x = x[1, l]\overline{x}[1, n-l]$ for every $l$, and blocks $B, B'$ resp. become $(n + 1 - B) = \{n + 1 - l : l \in B\}$ and $(n + 1 - B')$. Moreover, since upsteps become downsteps, and conversly, we get that $\mathrm{height}(x[1, l]) = \mathrm{height}(x[1, n]) + \mathrm{height}(\overline{x}[1, n - l])$. Therfore, $m$ is also the minimum of $\mathrm{height}(\overline{x}[1, l-1])$, where $l$ ranges over $(n + 1 - B)$ such that $\overline{x}_l$ is a downstep. Similarly, $m'$ is the minimum of $\mathrm{height}(\overline{x}[1, l])$, where $l$ ranges over $(n + 1 - B')$ such that $\overline{x}_l$ is an upstep.

After reading $B$, since $j$ is not yet matched, the stack necessarily contains an item corresponding to a word containing $j$; moreover, since all compressions in $B$ involve items with first letter in $B$, the first letter of that word is in $B$. From Fact 7, by the end of reading $B'$ that item has been discarded. Let $(h, \ell, f)$ be that item with its corresponding encoded subsequence $v$. Since the first letter $f$ of $v$ is in $B$, all of the letters of $v$ are in $B \cup B'$. By Fact 6, $v$ is a matching set, and, by Fact 8, its matching pairs in $B \times B'$ are all at different heights. So at height $d$, $v$ only contains $(j, j')$, which is not well-formed, plus possibly some pairs coming from $B \times B$ or from $B' \times B'$, pairs that are all well-formed by minimality of $i$. Overall at height $d$ the word $v$ is unbalanced, so by Fact 6, the probability that $v$ passes the hash test of **Algorithm 4** is at most $n^{-c}$, for a uniformly random choice of $\alpha$. So the algorithm is correct with probability $1 - n^{-c}$. $\qquad\square$

## 4 Lower bounds

We define a family of hard instances for $\textsc{Dyck}(2)$ as follows. For any word $Z \in \{a, b\}^n$, let $\overline{Z}$ be the matching word associated with $Z$. For given $m, n$, consider the following instances of length $\Theta(mn)$:

$$w = X_1 \overline{Y}_1 \overline{c}_1 c_1 Y_1 \ X_2 \overline{Y}_2 \overline{c}_2 c_2 Y_2 \ \ldots \ X_m \overline{Y}_m \overline{c}_m c_m Y_m \overline{X}_m \ \ldots \ \overline{X}_2 \ \overline{X}_1,$$

where for every $i$, $X_i \in \{0, 1\}^n$, $Y_i = X_i[n - k_i + 2, n]$ for some $k_i \in \{1, 2, \ldots, n\}$, and $c_i \in \{a, b\}$. The word $w$ is in $\textsc{Dyck}(2)$ if and only if, for every $i$, $c_i = X_i[n - k_i + 1]$.

Intuitively, for $m = n/\log n$ recognizing $w$ is difficult with space $o(n)$ because, after reading $X_i$, the streaming algorithm does not have enough space to store information about the bit at unknown index $(n - k_i + 1)$, so when it reads $c_i$ it is unable to decide whether $c_i = X_i[n - k_i + 1]$; and after reading $\overline{Y}_m$ it does not have enough space to store information about all indices $k_1, k_2, \ldots, k_m$, so when it reads $\overline{X}_m \ \ldots \ \overline{X}_2 \ \overline{X}_1$ it misses out on its second chance to check whether $c_i = X_i[n - k_i + 1]$ for every $i$. The proof contains some subtleties and is executed in the language of communication complexity.

We define a communication problem ASCENSION$(m)$ (Figure 5 in Appendix B) associated with the hard instances described above. For convenience, we replace suffixes by prefixes. Formally, in the problem ASCENSION$(m)$ there are $2m$ players $A_1, A_2, \ldots, A_m$ and $B_1, B_2, \ldots, B_m$. Player $A_i$ has $X_i \in \{0,1\}^n$, $B_i$ has $k_i \in [n]$, a bit $c_i$ and the prefix $X_i[1, k-1]$ of $X_i$. Let $\mathbf{X} = (X_1, X_2, \ldots, X_m)$, $\mathbf{k} = (k_1, k_2, \ldots, k_m)$ and $\mathbf{c} = (c_1, c_2, \ldots, c_m)$. The goal is to compute $f_m(\mathbf{X}, \mathbf{k}, \mathbf{c}) = \bigvee_{i=1}^m f(X_i, k_i, c_i) = \bigvee_{i=1}^m (X_i[k_i] \oplus c_i)$, which is 0 if $X_i[k_i] = c_i$ for all $i$, and 1 otherwise.

Motivated by the streaming model, we require each message to have length at most size bits, where the parameter size is a function of $m$ and $n$ and corresponds to the allowed space in the corresponding streaming algorithm. We also require the communication between the $2m$ participants in a one-pass protocol to be in the following order:

**Round 1**

For $i$ from 1 to $m - 1$, player $A_i$ sends message $M_{A_i}$ to $B_i$, then $B_i$ sends message $M_{B_i}$ to $A_{i+1}$; then $A_m$ sends message $M_{A_m}$ to $B_m$; then

**Round 2**

$B_m$ sends message $M_{B_m}$ to $A_m$; then
For $i$ from $m$ down to 2, $A_i$ sends a message $M'_{A_i}$ to $A_{i-1}$; then
$A_1$ computes the output.

To establish the hardness of solving ASCENSION$(m)$, we prove a *direct sum* result that captures its relationship to solving $m$ instances of a "primitive" problem MOUNTAIN defined as follows. In the problem MOUNTAIN (Figure 4 in Appendix B), Alice has an $n$-bit string $X \in \{0,1\}^n$, and Bob has an integer $k \in [n]$, a bit $c$ and the prefix $X[1, k-1]$ of $X$. The goal is to compute the Boolean function $f(X, k, c) = (X[k] \oplus c)$ which is 0 if $X[k] = c$, and 1 otherwise. In a one-pass protocol for MOUNTAIN, the communication occurs in the following order: Alice sends a message $M_A$ to Bob, Bob sends a message $M_B$ to Alice, then Alice outputs $f(X, k, c)$.

As mentioned in Section 1, we follow the "information cost" approach, a method that has been particularly successful in recent works on direct sum results. The method comes in a variety of flavours, each crafted to suit the application at hand. We describe the approach as adapted for ASCENSION$(m)$. Information cost is often defined in terms of the entire input and the full transcript of the protocol. We enforce both the nature of streaming algorithms and of our problem, by restricting our attention to only one message $M_{B_m}$ from the transcript. We also split the input in two parts, and measure the information in the message $M_{B_m}$ about one part $(\mathbf{k}, \mathbf{c})$, conditioned on the other part $\mathbf{X}$. In our case, the conditioning corresponds to information that is in the hands of the subsequent players. The closest such measures, of which we are aware, were considered in [17, 5].

The direct sum result is proven using the superadditivity of mutual information for inputs $(k_i, c_i)$ picked independently from a carefully chosen distribution. In the defining information cost, we measure mutual information with respect to a distribution on which the MOUNTAIN function is the constant 0, eventhough we consider protocols for the problem that are correct with high probability in the worst case (or, equivalently, when the inputs are chosen from a "hard" distribution). The use of this easy distribution collapses the function ASCENSION$(m)$ to an instance of MOUNTAIN in any chosen coordinate. We massage this (already established) technique into a form that is better suited to the streaming model and to proving lower bounds for the primitive function MOUNTAIN.

We finish by giving a combinatorial argument that protocols computing Mountain in the worst case necessarily reveal "a lot" of information even when its inputs are chosen according to the easy distribution. Privacy loss, a measure similar to information cost, has been studied previously in protocols for Index (see, e.g., [16, 14] and the references therein). Although this communication problem is closely related to Mountain, prior works study Index under hard distributions, and do not seem to extend directly to our case.

## 4.1 Information cost

We measure the *information cost* of a one-pass public-coin randomized protocol $P$ for Ascension$(m)$ (of the form described in the previous section), with respect to some distribution $\nu$ on the inputs $(\mathbf{X}, \mathbf{k}, \mathbf{c})$, by $\text{IC}_\nu(P) = \text{I}(\mathbf{k}, \mathbf{c} : M_{B_m} | \mathbf{X}, R)$, where $R$ denotes the public-coins of $P$. From this we define the *information cost* of the problem Ascension$(m)$ itself with respect to a distribution $\nu$ and error parameter $\delta$ as follows: $\text{IC}_\nu^{\text{pub}}(\text{Ascension}(m), \delta) = \min\left(\text{IC}_\nu(P)\right)$, where the minimum is over one-pass public-coin randomized protocols $P$ for the problem, with worst-case error at most $\delta$. Note that the information cost implicitly depends on the length size of each message.

For the problem Mountain we play a subtle game between public and private coins. We consider protocols in which Alice has access only to public coins $R$, whereas Bob additionally has access to some independent private coins $R_B$. We define $\text{IC}_\nu(P) = \text{I}(k, c : M_B | X, R)$, where $R$ denotes only the public-coins of $P$. Further, we define $\text{IC}_\nu^{\text{mix}}(\text{Mountain}, \delta) = \min\left(\text{IC}_\nu(P)\right)$, where $P$ ranges over "mixed" public and private coin randomized protocols with worst case error at most $\delta$ where Alice and Bob share public coins, and only Bob has access to extra private coins.

We also make use of a related measure of complexity for Mountain when $P$ ranges over protocols where Alice's message is deterministic, and Bob has access to private coins $R_B$: $\text{DIC}_\nu^{\text{mix}}(\text{Mountain}, \mu, \delta) = \min\left(\text{IC}_\nu(P)\right)$, i.e., the minimum information cost with respect to $\nu$, where $P$ ranges over protocols for Mountain, in which Alice's message $M_A$ is deterministic given her input $X$, while Bob may use his private coins $R_B$ to generate his message. Further, the distributional error of $P$ is at most $\delta$ when the inputs are chosen according to $\mu$. Note that in general, and certainly in our application, $\nu$ and $\mu$ may be different, meaning that we measure the information cost of the protocol with respect to some distribution $\nu$, while we measure its error under a potentially different distribution $\mu$. For later use, we recall that the distributional error under $\mu$ is $\text{Exp}_{(X,k,c)\sim\mu}\left(\Pr(P \text{ fails on } (X, k, c))\right)$, where the probability is over the private coins $R_B$ of Bob.

We begin by relating the information cost for protocols in which Alice is deterministic to that of mixed randomized protocols.

**Lemma 1.** $\text{DIC}_\nu^{\text{mix}}(\text{Mountain}, \mu, 2\delta) \leq 2 \times \text{IC}_\nu^{\text{mix}}(\text{Mountain}, \delta)$.

*Proof.* Consider a randomized protocol $P$ for Mountain with worst-case error at most $\delta$ such that $\text{IC}_\nu^{\text{mix}}(\text{Mountain}, \delta) = \text{IC}_\nu(P)$. We further assume that Alice and Bob have public coins $R$, and only Bob has extra private coins $R_B$. Then

$$\text{IC}_\nu^{\text{mix}}(\text{Mountain}, \delta) = \text{Exp}_r\left(\text{I}(k, c : M_{B_m} | X, R = r)\right),$$

Since $P$ has worst-case error at most $\delta$, it has distributional error at most $\delta$ under $\mu$:

$$\text{Exp}_r\left(\text{Exp}_{(X,k,c)\sim\mu}\left(\Pr(P \text{ fails on } (X, k, c) | R = r)\right)\right) \leq \delta.$$

Therefore, by the Markov inequality, there is a set $\mathcal{R}$ with $\Pr(r \in \mathcal{R}) \geq \frac{1}{2}$ such that

$$\forall r \in \mathcal{R} \qquad \underset{(X,k,c)\sim\mu}{\mathrm{Exp}} \big( \Pr(P \text{ fails on } (X,k,c)|R=r) \big) \quad \leq \quad 2\delta.$$

Now consider the information cost of $P$ under the distribution $\nu$ over inputs. We have

$$\underset{r \in \mathcal{R}}{\mathrm{Exp}} \big( \mathrm{I}(k,c : M_{B_m}|X, R=r) \big) \quad \leq \quad 2 \times \mathrm{IC}_\nu^{\mathrm{mix}}(\textsc{Mountain}, \delta),$$

since $\mathcal{R}$ has probability mass at least $1/2$. Therefore, there exists an $r \in \mathcal{R}$ such that $\mathrm{I}(k,c : M_{B_m}|X, R = r) \leq 2 \times \mathrm{IC}_\nu^{\mathrm{mix}}(\textsc{Mountain}, \delta)$. Let $P_r$ be the protocol obtained by fixing the public coins used in $P$ to $r$. Then Alice's message $M_A$ is deterministic. By definition of $\mathcal{R}$, the protocol $P_r$ has distributional error at most $2\delta$ under $\mu$, and $\mathrm{IC}_\nu(P) \leq 2 \times \mathrm{IC}_\nu^{\mathrm{mix}}(\textsc{Mountain}, \delta)$. $\qquad\square$

## 4.2 Information cost of Mountain

As explained beore, and formally proved in the next section, the information cost approach entails showing that the Mountain problem is "hard" even when we restrict our attention to an easy distribution. We prove such a result here.

Let $\mu$ be the distribution over inputs $(X,k,c)$ in which $X$ is a uniformly random $n$-bit string, $k$ is a uniformly random integer in $[n]$ and $c$ a uniformly random bit. This is a hard distribution for Mountain (as is implicit in [20, 3]). We consider information cost of Mountain under the distribution $\mu_0$ obtained by conditioning $\mu$ on the event that the function value is 0: $\mu_0(X,k,c) = \mu(X,k,c|X[k]=c)$.

**Lemma 2.** *If* size $\leq n/100$, *then* $\mathrm{DIC}_{\mu_0}^{\mathrm{mix}}(\textsc{Mountain}, \mu, 1/16n^2) = \Omega(\log n)$.

*Proof.* Let $P$ be a randomized protocol for Mountain, where Alice's message $M_A$ is deterministic, with distributional error at most $1/16n^2$ under the distribution $\mu$, such that $|M_A| \leq n/100$. We prove that $\mathrm{IC}_{\mu_0}(P) = \Omega(\log n)$. In the following, all expressions involving mutual information and entropy are with respect to the distribution $\mu_0$.

By Markov inequality, there are at least $2^{n-1}$ strings $U$ on which $P$ fails with error at most $1/8n^2$ on average on input $(U,k,c)$, where $(k,c)$ varies uniformly. Let $S \subseteq \{0,1\}^n$ of size at least $2^{n-1}$ be the set of such strings $U$. Then $P$ has error probability less than $1/4n$ on input $(U,k,c)$, for every pair $(k,c)$.

Let $\alpha$ be a possible message $M_A$ from Alice to Bob when her inputs range in $S$, and let $S_\alpha = \{U \in S : M_A(U) = \alpha\}$. For every string $V \in S_\alpha$, we bound from below the mutual information of $k$ and $M_B$, the randomized message that Bob sends back to Alice, as $k$ varies. For this we construct a set $I \subseteq [n]$ such that the message distributions $m_k = M_B(\alpha, V[1,k-1], k, V[k])$ for $k \in I$ are pairwise well-separated in $\ell_1$ distance. This is in turn established by exhibiting, for each $k \in I$, a string $V_k \in S_\alpha$ such that $V[1,k-1] = V_k[1,k-1]$ and $V[k] \neq V_k[k]$. The details follow.

Associate with $S_\alpha$ its dictionary $T$, a 2-rank tree (a tree with either 1 or 2 children at any internal node), all whose nodes except the root are labeled by bits; the root has an empty label. Each string $V$ in $S_\alpha$ is in one-to-one correspondence with a top-down path $\pi$ in $T$ from the root to one of its leaves, where the label of the $(i+1)$th node in $\pi$ is $V[i]$. We identify $V \in S_\alpha$ with the path $\pi$ in $T$, and refer to this path as $V$.

The tree $T$ has $|S_\alpha|$ leaves, each at depth $n$. For a fixed $V \in S_\alpha$, let $I$ be the set of integers $k$ such that the $(k+1)$th node in path $V$ has out-degree 2. By construction, for every $k \in I$ there exists another string, say, $V_k \in S_\alpha$ such that $V[1,k-1] = V_k[1,k-1]$ and $V[k] \neq V_k[k]$.

Set $c_k = V[k]$ for every $k \in [n]$. Then the message distributions satisfy $M_B(\alpha, V[1,k-1], k, c_k) = M_B(\alpha, V_k[1,k-1], k, c_k)$, for all $k \in I$. Let $m_k = M_B(\alpha, V[1,k-1], k, c_k)$. Let $k, k' \in I$ be distinct indices such that $k < k'$. As $V_{k'}[1, k'-1] = V[1, k'-1]$, the message distribution $M_B(\alpha, V_{k'}[1, k-1], k, c_k)$ on input $(V_{k'}, k, c_k)$ equals $m_k$, and also $M_B(\alpha, V_{k'}[1, k'-1], k', c_{k'})$ on input $(V_{k'}, k', c_{k'})$ equals $m_{k'}$. However, $V_{k'}[k] = V[k] = c_k$, so the function evaluates to 0 on input $(V_{k'}, k, c_k)$, and $V_{k'}[k'] \neq V[k'] = c_{k'}$, so the function value is 1 on $(V_{k'}, k', c_{k'})$. The protocol $P$ computes its outputs from $m_k, V_{k'}$ and $m_{k'}, V_{k'}$, respectively, on these instances, and errs with probability at most $1/4n$.

We use the above property of the distributions to lower bound the mutual information of $k$ in the message $M_B$, given $V$.

**Proposition 2.** $\mathrm{I}(k : M_B | X = V) \quad \geq \quad \left( \dfrac{4|I| - n}{4n} \right) \log n - 2.$

(We prove this below.)

Next, we observe from the properties of 2-rank trees that the number of strings $V \in S_\alpha$ for which $|I| = l$ is at most $2^l$. The number of $V$ for which $|I| \leq l - 2$ is therefore at most $2^{l-1}$. Now fix $l = \log |S_\alpha|$, and note that the proportion of $V \in S_\alpha$ with $|I| \geq l - 1$ is at least $1/2$. Therefore $\mathrm{Exp}_{V \in S_\alpha} |I| \quad \geq \quad \frac{l-1}{2}$.

We now concentrate on the messages $\alpha$ such that $\Pr_{X \text{ uniform}}(M_A(X) = \alpha | X \in S) \geq 2^{-n/10}$. Then $l = \log |S_\alpha| \geq n - 1 - n/10 = 0.9n - 1$, and for $n$ large enough, $\mathrm{Exp}_{V \in S_\alpha} \mathrm{H}(k | M_B, X = V) \leq 2 + \frac{9}{10} \log n$, and therefore $\mathrm{Exp}_{V \in S_\alpha}(\mathrm{I}(k, c : M_B | X = V)) \quad \geq \quad \frac{1}{10} \log n - 2$.

The net probability of messages $\alpha$ which have probability at most $2^{-n/10}$ given $X \in S$ is at most $2^{n/100} 2^{-n/10} = 2^{-9n/10}$, which is negligible. Therefore we conclude that $\mathrm{I}(k, c : M_B | X) = \Omega(\log n)$. $\quad \square$

*Proof of Fact 2.* Fix a string $V$, and the corresponding set of indices $I$. Suppose we are given as input a distribution $m = m_k$, for some $k \in I$. We recover $k$ using the following procedure $\Pi$:

1. For each $k' \in I$, simulate the Alice's computation of the output in the protocol $P$, by setting $M_B = m$, the input distribution, and $X = V_{k'}$.
2. Let $(d_k)_{k' \in I}$ be the sequence of outputs Alice generates from the above simulation. Output the largest $k'$ for which $d_{k'} = 1$. This is our guess for $k$.

On input $m_k$, the simulation of $P$ above generates $d_k = 1$, and $d_{k'} = 0$ for $k' > k$, with probability at least $1 - 1/4n$ for any fixed $k' \geq k$. Therefore, the procedure outputs $k' = k$ with probablity at least $3/4$.

We now argue that the entropy of $k$ is significantly reduced when given $M_B, X$, under the distribution $\mu_0$ (i.e., when $c_k = X[k]$). This is equivalent to saying that the mutual information of $k$ and $M_B$ is high. When the inputs are picked according to the distribution $\mu_0$, we have

$$\mathrm{I}(k, c : M_B | X = V) \quad = \quad \mathrm{H}(k | X = V) - \mathrm{H}(k | M_B, X = V) \quad = \quad \log n - \mathrm{H}(k | M_B, X = V).$$

We bound from above the conditional entropy $\mathrm{H}(k | M_B, X = V)$. We first separate the values of $k \notin I$ as follows. Let $p = |I|/n$, and define the Boolean random variable $J$ as 1 iff $k \in I$. We have

$$
\begin{aligned}
\mathrm{H}(k | M_B, X = V) \quad &= \quad \mathrm{H}(kJ | M_B, X = V) \\
&= \quad \mathrm{H}(J | M_B, X = V) + \mathrm{H}(k | M_B, X = V, J) \\
&= \quad \mathrm{H}(p) + (1 - p)\mathrm{H}(k | M_B, X = V, k \notin I) + p\,\mathrm{H}(k | M_B, X = V, k \in I) \\
&\leq \quad 1 + (1 - p)\log n + \mathrm{H}(k | M_B, X = V, k \in I) \\
&\leq \quad 1 + (1 - p)\log n + \mathrm{H}(k | K, X = V, k \in I),
\end{aligned}
$$

where $K$ is the random variable computed by our finding procedure $\Pi$, and the final step follows from the Data Processing Inequality. For any fixed $k \in I$, given $M_B$ the procedure $\Pi$ computes $K = k$ with probability at least $3/4$. By the Fano Inequality, we have

$$\mathrm{H}(k | K, X = V, k \in I) \quad \leq \quad \mathrm{H}(\tfrac{1}{4}) + \tfrac{1}{4}\log(|I| - 1) \quad \leq \quad 1 + \tfrac{1}{4}\log n.$$

$\square$

By combining Lemmas 1 and 2 we get

**Theorem 3.** $\mathrm{IC}^{\mathrm{mix}}_{\mu_0}(\text{MOUNTAIN}, 1/42n^2) = \Omega(\log n)$.

### 4.3 Reduction to from ASCENSION$(m)$ to MOUNTAIN

We now study the information cost of ASCENSION$(m)$ for the distribution $\mu_0^m$ over $(\{0,1\}^n \times [n] \times \{0,1\})^m$ for $\mathbf{X} = (X_1, X_2, \ldots, X_m)$, $\mathbf{k} = (k_1, k_2, \ldots, k_m)$ and $\mathbf{c} = (c_1, c_2, \ldots, c_n)$. We state a direct sum property that relates this cost to that of one instance of MOUNTAIN, and then conclude.

**Lemma 3.** $\mathrm{IC}^{\mathrm{pub}}_{\mu_0^m}(\text{ASCENSION}(m), \delta) \geq m \times \mathrm{IC}^{\mathrm{mix}}_{\mu_0}(\text{MOUNTAIN}, \delta)$.

*Proof.* Let $P$ be a public-coin randomized protocol for ASCENSION$(m)$ with worst-case error $\delta$ such that $\mathrm{IC}_{\mu_0^m}(P) = \mathrm{IC}^{\mathrm{pub}}_{\mu_0^m}(\text{ASCENSION}(m), \delta)$.

From $P$, we construct the following protocol $P'_j$ for MOUNTAIN, where $j \in [n]$. Let $(X, k, c)$ be the input for MOUNTAIN.

1. Alice sets $A_j$'s input $X_j$ to its input $X$.
2. Bob sets $B_j$'s input $(k_j, X_j[1, k_j - 1], c_j)$ to its input $(k, X[1, k-1], c)$.
3. Alice and Bob generate, using public coins, $(X_i, k_i, c_i)$ according to $\mu_0$, independently for all $i < j$, and $X_i$ uniformly independently for $i > j$.
4. Bob generates $(k_i)$ uniformly independently for $i > j$, but using his private coins. Then Bob sets $c_i = X_i[k_i]$ for $i > j$ (so that $(X_i, k_i, c_i)$ are now according to $\mu_0$, independently for all $i < j$).
5. Alice and Bob run the protocol $P$ by simulating the players $(A_i, B_i)_{i=1}^m$ as follows:
   (a) Alice runs $P$ until she generates the message $M_{A_j}$ from player $A_j$. She sends this message to Bob.
   (b) Bob continues running $P$ until he generates the message $M_{B_m}$ from player $B_m$. He sends this message to Alice.
   (c) Alice completes the rest of the protocol $P$ until the end, and produces as output for $P'_j$, the output of player $A_1$ in $P$.

By definition of the distribution $\mu_0$, we have $f(X_i, k_i, c_i) = 0$ for all $i \neq j$. So $f_m(\mathbf{X}, \mathbf{k}, \mathbf{c}) = f(X, k, c)$, and each protocol $P'_j$ computes the function $f$, i.e., solves MOUNTAIN, with worst-case error $\delta$.

We prove that $\mathrm{IC}_{\mu_0^m}(P) = \sum_j \mathrm{IC}_{\mu_0}(P'_j)$, which implies the result, since only Bob uses private coins in $P'_j$.

Let $R$ denote the public coins used in the protocol $P$. By applying the chain rule to $\mathrm{IC}_{\mu_0^m}(P)$, we get

$$\begin{aligned} \mathrm{IC}_{\mu_0^m}(P) &= \mathrm{I}(\mathbf{k}, \mathbf{c} : M_{B_m} | \mathbf{X}, R) \\ &= \sum_j \mathrm{I}(k_j, c_j : M_{B_m} | \mathbf{X}, k_1, c_1, \ldots, k_{j-1}, c_{j-1}, R) \end{aligned}$$

Let $R_j = (R, (X_i)_{j \neq i}, (k_i, c_i)_{i<j})$. These are all the public random coins used in the protocol $P'_j$, and any further random coins $(k_i, c_i)_{i>j}$ are used only by Bob. Since for all $j$

$$\mathrm{IC}_{\mu_0}(P'_j) = \mathrm{I}(k_j, c_j : M_{B_m} | X_j, R_j),$$

which is the same as $\mathrm{I}(k_j, c_j : M_{B_m} | \mathbf{X}, k_1, c_1, \ldots, k_{j-1}, c_{j-1}, R)$, the direct sum result follows. $\square$

**Theorem 4.** *Let $P$ be a public-coin randomized protocol for* Ascension$(n/\log n)$ *with worst-case error probability $1/42n^2$, then* size $= \Omega(n)$.

*Proof.* Let $m = n/\log n$ and $\delta = 1/42n^2$, and let $P$ be a public-coin randomized protocol for Ascension$(m)$ with worst-case error probability $\delta$. Obviously, $\mathrm{IC}_{\mu_0^m}(P)$ is at most size. On the other hand, by definition $\mathrm{IC}_{\mu_0^m}^{\mathrm{pub}}(\textsc{Ascension}(m), \delta)$ is less than or equal to $\mathrm{IC}_{\mu_0^m}(P)$. By Lemma 3, we have $\mathrm{IC}_{\mu_0^m}^{\mathrm{pub}}(\textsc{Ascension}(m), \delta) \geq m \times \mathrm{IC}_{\mu_0}^{\mathrm{mix}}(\textsc{Mountain}, \delta)$. By Theorem 3, we get $\mathrm{IC}_{\mu_0}^{\mathrm{mix}}(\textsc{Mountain}, \delta) = \Omega(\log n)$. Combining yields size $= \Omega(m \log n) = \Omega(n)$. $\qquad\square$

**Corollary 1.** *Every one-pass randomized streaming algorithm for the matching parenthesis problem of words of length $n'$ with (two-sided) error $\mathrm{O}(1/n' \log n')$ uses $\Omega(\sqrt{n' \log n'})$ space.*

*Proof.* Assume we have a one-pass randomized streaming algorithm for the matching parenthesis problem of words of length $n'$ with (two-sided) error $\mathrm{O}(1/n' \log n')$ uses $\Omega(\sqrt{n' \log n'})$ space. Then, by the discussion at the beginning of Section 4, there is a public-coin randomized protocol for Ascension$(n/\log n)$ with $n = \Theta(\sqrt{n' \log n'})$ and with worst-case error probability $1/42n^2$. By Theorem 4, the messages must have size $\Omega(n)$, and therefore, by the discussion, the streaming algorithm must have space $\Omega(n) = \Omega\sqrt{n' \log n'}$. $\qquad\square$

## Acknowledgements

F.M. would like to thank earlier discussions with Michel de Rougemont, Miklos Santha, Umesh Vazirani, and especially with Pranab Sen, who, among other things, noticed that the logarithmic space algorithm for Identity$(s)$ in [18] can be converted to a one-pass randomized streaming algorithm with logarithmic space.

## A  Example of execution of Algorithms 1 and 3

Figure 1 shows an example of execution of our one-pass algorithm. Here there are eight blocks, and they are shown after the internal simplifications have already been done. The dotted vertical lines mark times at which the stack changes size, either starting a new stack item (for example, at time $t_0$) or discarding a stack item (for example, at time $t_4$). Note that blocks and stack items are staggered: the first item incorporates the first block and the downsteps of the second block, the second item incorporates the upsteps of the second block and the downsteps of the third block, etc. The bullets mark times when the algorithm checks and discards an item. The horizontal lines go from the time when a stack item is created to the time when it is checked and discarded. For example, at time $t_7$ the algorithm checks and discards an item $(h_m, \ell_m), (d, \ell_d)$ such that $h_m$ incorporated the upsteps marked in bold on the figure, namely $x(t_1, t_2]$, and $d$ incorporated the downsteps marked in bold on the figure, namely $x(t_2, t_3]$, $x(t_4, t_5]$ and $x(t_6, t_7]$.

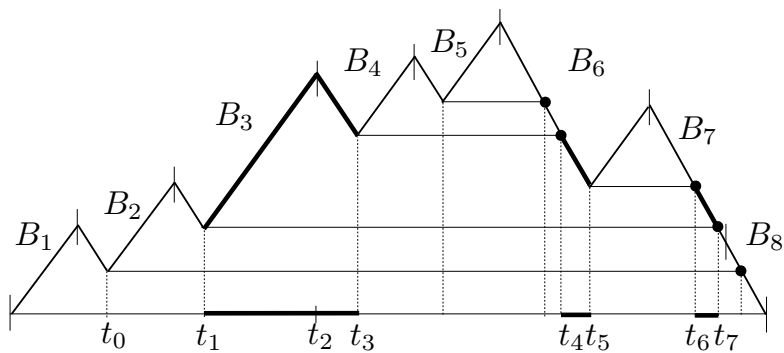

Figure 1: Example of execution of Algorithm 1

Figure 2 illustrates the logarithmic block decomposition of the input word into all the blocks that will be activated during one-pass. They are identical from the left-to-right pass and the right-to-left pass since thanks to padding the input length is a power of 2. At every instant, only one $i$-block is activated for each $i$.

Figure 3 gives an intuition of the proof of Fact 7. The bold-face lines represent matching pairs between the two $(i-1)$-blocks $B, B'$ within the same $i$-block $B_i$. In the case of the figure, those pairs are checked during the left-to-right pass, since the minimum height $m$ within the left $(i-1)$-block $B$ is larger than the minimum height $m'$ with the right $(i-1)$-block $B'$ (during the right-to-left pass, they are compressed without any checks when $B_i$ is processed).

## B  Figures for MOUNTAIN and ASCENSION$(m)$

Figure 4 presents an input stream with its division between players Alice and Bob. The horizontal axis represents the length of the stream seen so far, and the vertical axis represents the corresponding height. We introduce a potential mismatch denoted by letter $c$ in Bob's input.

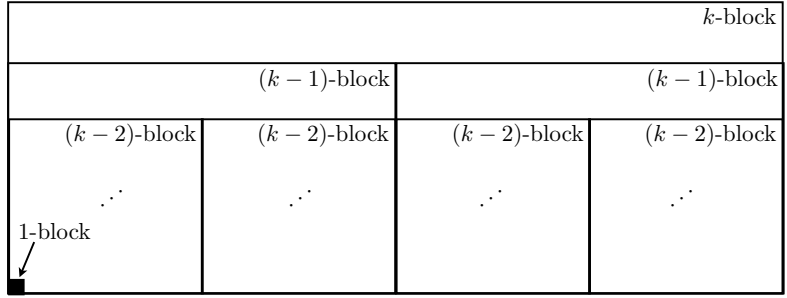

Figure 2: Decomposition in block-structure

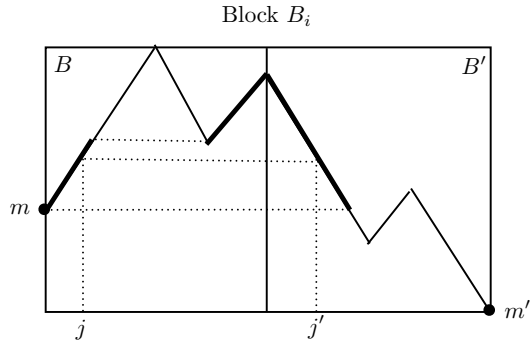

Figure 3: Illustration of Fact 7 for an example of execution of Algorithm 3

Figure 5 presents the $m$-fold nesting of the above stream. The stream is now divided between $2m$ players. There are $m$ potential mismatches each caused by the letter $c_i$ in $B_i$'s input.

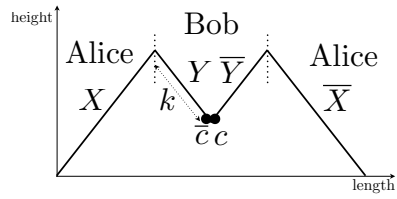

Figure 4: Problem MOUNTAIN: $\overline{Y}[1, k-1] = \overline{X}[1, k-1]$. The word is well-formed if and only if $c = \overline{X}[k]$.

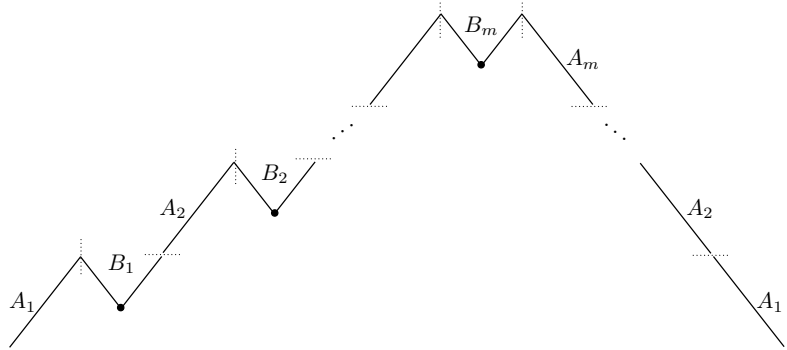

Figure 5: Problem ASCENSION($m$): The word is well-formed if and only $c_i = \overline{X}_i[k_i]$, for all $i$.